\documentclass[twocolumn,showpacs,%
  nofootinbib,aps,superscriptaddress,%
  eqsecnum,prd,notitlepage,showkeys,10pt]{revtex4-1}

\usepackage{amssymb}
\usepackage{xcolor}
\usepackage{amsmath}
\usepackage{graphicx, wrapfig, lipsum, caption}
\usepackage{subfigure}
\usepackage{dcolumn}
\usepackage{hyperref}
\usepackage{blindtext}
\usepackage{setspace}
\usepackage{booktabs}
\usepackage{romannum}

\begin{document}

\title{Automatic Estimation of Pedestrian Gait Features using a single camera recording: Algorithm and Statistical Analysis for Gender Difference and Obstacle Interactions}

\author{Kanika Jain$^2$, Abhishek Gupta$^3$\thanks{some info}, Indranil Saha Dalal$^{2*}$, Anurag Tripathi$^{2*}$, Shankar Prawesh$^{*}$}

\affiliation{Industrial \& Management Engineering, Indian Institute of Technology Kanpur, India\\$^2$Department of Chemical Engineering, Indian Institute of Technology Kanpur, India\\\begin{minipage}{0.68\textwidth}$^3$Optum Global Solutions Pvt. Ltd., India \begin{scriptsize}\\(Disclaimer: The research undertaken and the views thus expressed are in the personal capacity of the author, and does not represent the research interest or views or opinions of the organization, nor does it constitute an endorsement or affiliation. Reader's discretion is advised.)\end{scriptsize}\end{minipage}}

\begin{abstract}
\textbf{Abstract:} The pedestrian gait features - body sway frequency, amplitude, stride length, and speed, along with pedestrian personal space and directional bias, are important parameters to be used in different pedestrian dynamics studies. Gait feature measurements are paramount for wide-ranging applications, varying from the medical field to the design of bridges. Personal space and choice of direction (directional bias) play important role during crowd simulations. In this study, we formulate an automatic algorithm for calculating the gait features of a trajectory extracted from video recorded using a single camera attached on the roof of a building. Our findings indicate that females have 28.64\% smaller sway amplitudes, 8.68\% smaller stride lengths, and 8.14\% slower speeds compared to males, with no significant difference in frequency. However, according to further investigation, our study reveals that the body parameters are the main variables that dominate gait features rather than gender. We have conducted three experiments in which the volunteers are walking towards the destination a) without any obstruction, b) with a stationary non-living obstacle present in the middle of the path, and c) with a human being standing in the middle of the path. From a comprehensive statistical analysis, key observations include no significant difference in gait features with respect to gender, no significant difference in gait features in the absence or presence of an obstacle, pedestrians treating stationary human beings and stationary obstacles the same given that the gender is same to match the comfort level, and a directional bias towards the left direction, likely influenced by India’s left-hand traffic rule.
\end{abstract}

\maketitle

\section{Introduction} \label{sec:Intro}

\begin{figure}
    
    \centering
    \includegraphics[width= 1\linewidth]{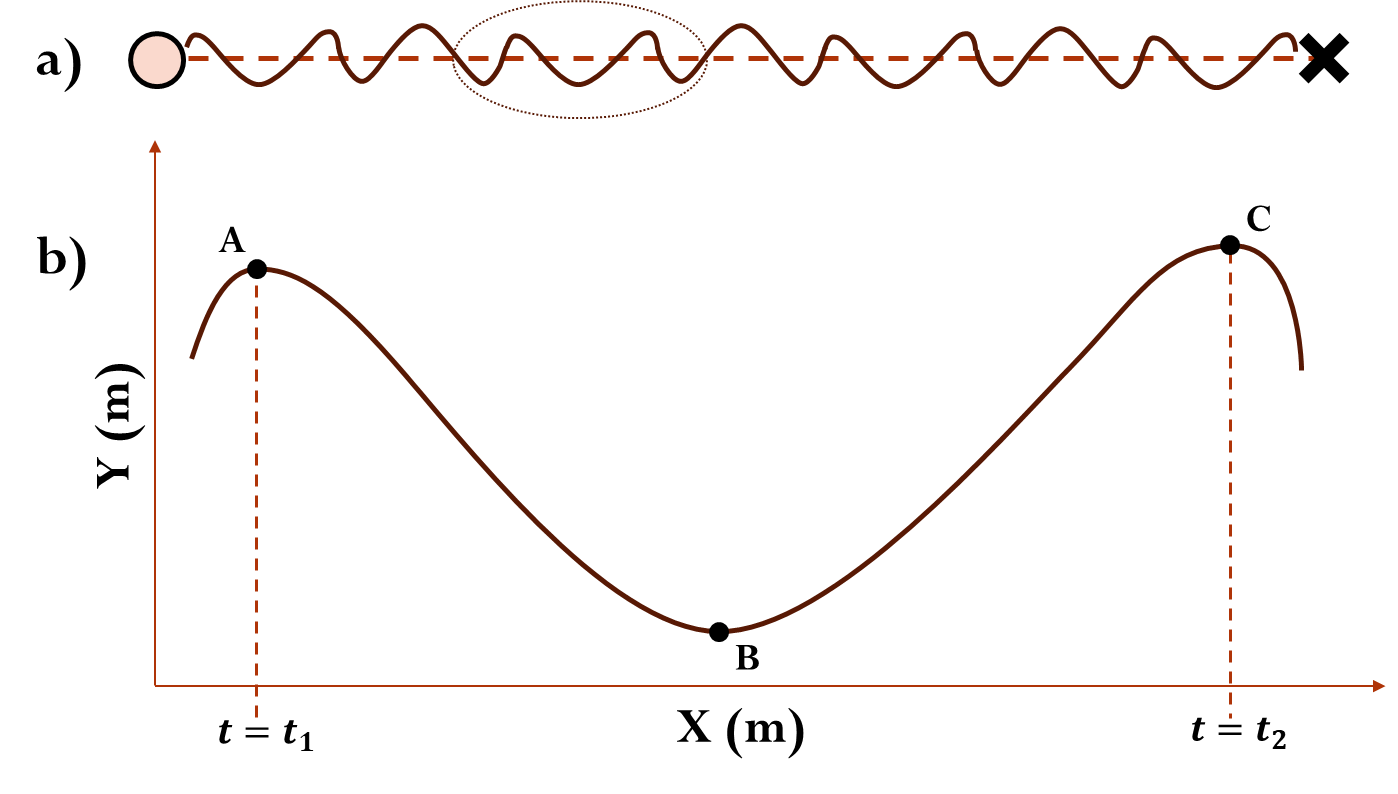}
    \caption{{\footnotesize The figure a) shows an animated representation of a pedestrian's trajectory walking towards the goal in a cyclic motion rather than in a straight line. The figure b) contains zoomed version of a single cycle, termed as `gait cycle' containing peaks and valleys representing two consecutive steps where the first step starts with peak A at time $t_1$ and the second step ends with peak C at time $t_2$.}}
    \label{fig:gaitCycle}
\end{figure}

The practice of walking in humans may be characterized as a gait. The Oxford Dictionary has a more common interpretation, defining gait as a `way of walking'. However, in the literature, the term `gait' often refers to the manner or style of locomotion \cite{racic2009experimental}. The motion of an individual on foot is a repetitive occurrence characterized by the completion of a single gait cycle, which involves a series of two consecutive steps \cite{perry2024gait}. A pedestrian moving in a straight line is expected to have a straight-line trajectory, but in reality, the pedestrian walks in a cyclic motion in order to balance the body weight while walking, as represented in Figure \ref{fig:gaitCycle}. As mentioned, the term `gait' refers to the way of walking and a single cycle in this cyclic motion is termed as `gait cycle'. The gait cycle will have multiple gait features, for instance, Jang identified 26 parameters for gender detection, subsequently narrowing them down to 19 through feature selection \cite{yoo2005gender}. In our study, the gait features have four key components: body sway frequency \cite{jia2019experimental, zhang2018height}, amplitude \cite{jia2019experimental, wang2010body}, stride length \cite{sekiya1997optimal, yang2022real, zhang2018height}, and speed. Body sway refers to the subtle postural adjustments for balance, with frequency calculated as $1/(t_2-t_1)$ in a single gait cycle from Figure \ref{fig:gaitCycle} b). Amplitude is half of the vertical length between points A and B. Stride length is the direct distance covered in two consecutive steps. In the figure, the stride length is the horizontal distance between points A and C. The speed is simply the direct distance traveled per unit of time, hence, the horizontal distance between points A and C divided by $(t_2-t_1)$. In addition to these gait features, we have also calculated the personal space maintained by the volunteers while walking around an obstacle and the directional bias observed passing it.

The study of gait features has been previously performed in the literature because of its wide-ranging applications, varying from the medical field to the design of bridges. The understanding of gait characteristics is valuable in several medical domains, such as the examination of child development \cite{thevenon2015collection}, the assessment of balance in elderly people \cite{wang2010body}, ambulatory diagnostics, body functioning, rehabilitation \cite{sekiya1997optimal, cai2017single}, diseases \cite{chatterjee2020analysis, bartsch2007fluctuation}, etc. The generation of vibrations resulting from the lateral movement of pedestrians is a crucial factor in the designing process of a bridge. A comprehensive investigation of the interaction between pedestrians and bridges, specifically focusing on the body sway motion of individuals, is available in literature \cite{abdulrehem2009low, belykh2016bistable}. Various methods have been suggested in the literature to capture gait features. For instance, one approach involves manual measurement by applying ink spots on the heels while walking \cite{sekiya1997optimal}.  Another method entails the placement of sensors on both feet \cite{yang2022real}. Several studies have also used accelerometers to estimate the gait features \cite{renaudin2012step, zhang2018height, alvarez2006comparison}. These estimations are conducted through a range of methodologies, including the Inverted Pendulum Model \cite{alvarez2006comparison, belykh2016bistable, h2015lateral}, the method based on the peak values and the valley values of the acceleration in the center of gravity \cite{weinberg2002using}, and the method based on linear combination \cite{shin2011adaptive}, among others. Recent studies have also been estimating the features by reading the sensor data collection with the help of smartphones\cite{yao2020robust, sadhukhan2023irt}. Relatively fewer studies have used video recording to measure the gait features \cite{wang2010body, wang2018linking}. J. Wang et al. utilized PeTrack \cite{wang2018linking}, while F. Wang et al. employed the Camera Calibration Toolbox to generate voxel persons \cite{wang2010body}. In this study, a single camera is used to capture visual data from the roof of a 25 m high building. This simple method is preferred due to the ease of data collection with minimum hindrance or comfort of participants, making it versatile for usage in general public.

The trajectories of participants are extracted to calculate various gait features and the effect of the presence of other obstacles and pedestrians on them. Note that the impact of obstacles on crowd dynamics has been extensively documented in the literature \cite{severiukhina2017study, karbovskii2019impact, chen2024experimental, shiwakoti2019review}. For instance, the formation of self-organized lanes in a dense crowd flow has been observed \cite{lian2015experimental}. Additionally, the presence of obstacles has been found to facilitate efficient outflow of pedestrians during evacuation or exit \cite{yanagisawa2009introduction, yanagisawa2010study}. The literature suggests that the behavior of a pedestrian is influenced by various factors such as cultural background \cite{subaih2020experimental}, age \cite{zhang2016homogeneity}, gender \cite{cao2018stepping}, and body mass index (BMI) \cite{aghabayk2021investigation}. Such behavioral alterations include effects on various gait features such as walking speeds \cite{aghabayk2021investigation, subaih2020experimental}, directional choice, personal space, density \cite{cao2018stepping}, and therefore, fundamental diagram \cite{seyfried2005fundamental}. The microscopic-level analysis of pedestrian behavior in response to obstacles has been subject to investigation. Several classical models, including the cellular automata model \cite{burstedde2001simulation}, social force model \cite{helbing1995social, hu2023anticipation}, and velocity-based model \cite{fiorini1998motion, hu2022effects}, have undergone modifications to incorporate obstacle-evading behavior. The effect of obstacles and other pedestrians on the gait features has been relatively less explored. In this study, we performed controlled experiments aiming at the analysis of behavioral changes in the volunteers while passing a stationary, non-living-human-sized obstacle in comparison to a stationary human being. The personal gap, as defined in this context, pertains to the lateral distance between the pedestrian and the obstacle. The study examines the impact of obstacles on the personal gap and the directional choices made by pedestrians, specifically in the context of Indian pedestrians. 

\begin{figure}

    \centering
    \includegraphics[width= 1\linewidth]{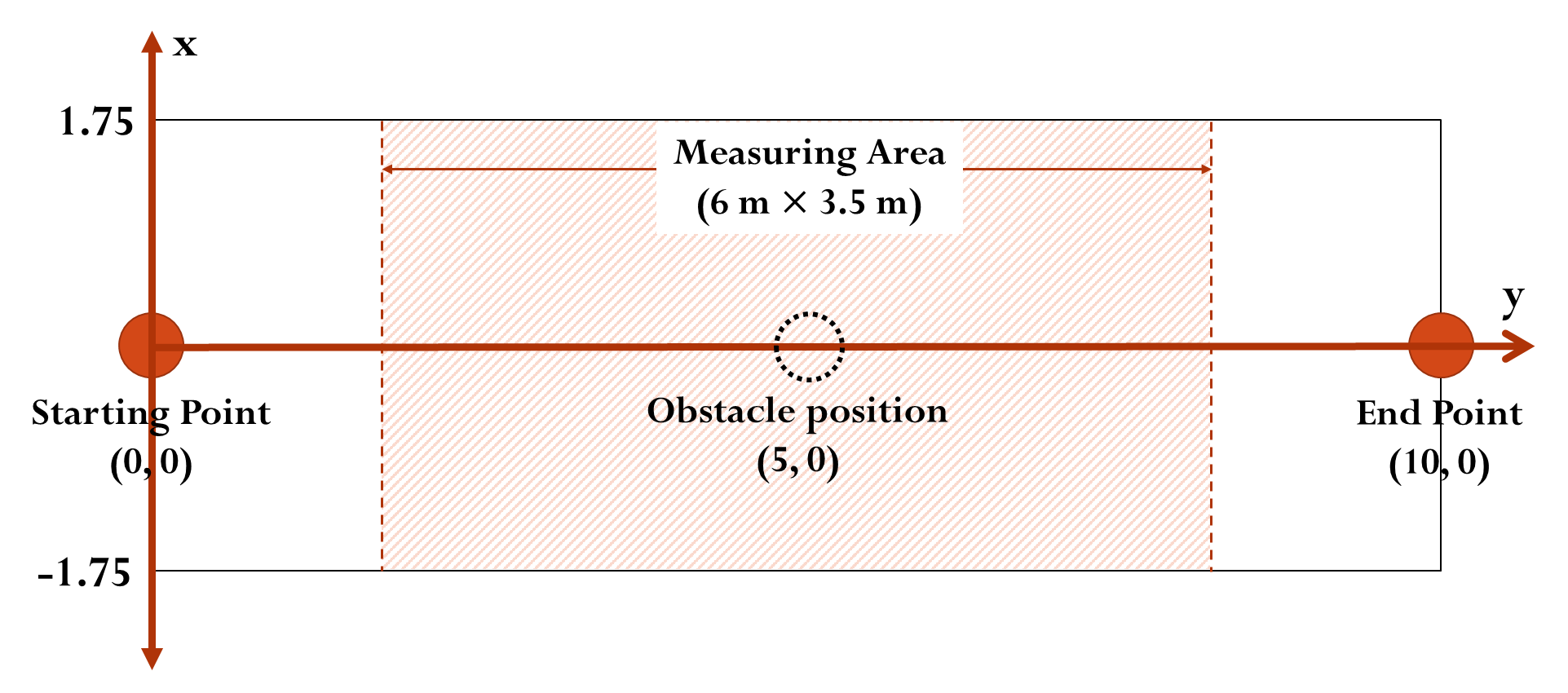}
    \caption{{\footnotesize Schematic for three experimental setups. Experiment 1: No obstacle is present in between the path. Experiment 2: A non-living human-sized, stationary obstacle is present at the obstacle position. Experiment 3: A person is standing at the obstacle position. For more details, see the text.}}
    \label{fig:ExpSetup}
\end{figure}

In this study, we have extensively studied the gait features of pedestrians in the Indian scenario across gender differences. Toward this, we have developed an automatic algorithm (explained later in Figure \ref{fig:GFflowchart}) that can calculate these features of a trajectory extracted from a video shot. A detailed video explaining the same is provided at \url{https://github.com/kanika201293/Gait-Feature-Calculation/issues/1}. In addition to gait features, the personal space maintained between pedestrians and/or obstacles is also calculated across genders, along with a quantifying bias in selecting directions while walking. We observe and discuss the similarities and differences of the trends and mean value of features with those reported from some earlier studies, for volunteers of another country. To the best of our knowledge, no equivalent study exists for the Indian scenario, especially analyzing the gender differences. Such investigations are useful for designing of public spaces, where pedestrians of both genders interact while walking.

\section{Experimental Setup} \label{sec:ExpSetup}

\begin{table} [h]
\centering

\def\arraystretch{1.3}

\scriptsize

\caption{{\footnotesize This table presents the distribution of total volunteers by gender along with key body parameters. The median and interquartile range (IQR) for shoulder length and height are provided separately for male and female participants.}}
\label{table:voluteers}
\begin{tabular}{|c|c|c|}
\hline
 & \textbf{Males} & \textbf{Females}\\\hline
\textbf{Volunteers} & 41 & 33\\\hline
\textbf{Shoulder} & median = 47 & median = 39\\
\textbf{Length (cm)} & IQR = 44.75 - 49 & IQR = 37.75 - 40\\\hline
\textbf{Height (cm)} & median = 170 & median = 152\\
 & IQR = 165.75 - 176.5 & IQR = 145.75 - 156.25\\\hline
\end{tabular}

\end{table} 

The experiments were conducted with volunteers of both genders on the IIT Kanpur campus. The video footage was captured from the roof of a 25 m high building. The trajectories were recorded using a Realme5pro mobile camera at $2\times$ magnification. The resolution of the images is 1920×1080 pixels, and the videos are recorded at 30 frames per second (fps). The experimental setup is depicted in Figure \ref{fig:ExpSetup}, where the main experimental area is a $10\;m$ long and $3.5\;m$ wide rectangle. Two red/filled circles indicate the starting and the end points of the volunteers, while the dashed circle at the center indicates the position of the obstacle. All the analyses reported were performed within the designated measurement area of $6m\times3.5m$, shown as the shaded region in Figure \ref{fig:ExpSetup}. The trajectory data in the gap of $2\;m$ on the left and right of the measurement region is not included to extract gait features to avoid entry and exit effects.

The study comprises three different experiments. In each experiment, the volunteers walk from the starting point to the end point, and their video during the walk is recorded.\\

\begin{itemize}
    \item Experiment 1 is performed in the absence of any obstacle in the path of the participants to examine different gait features during normal walking.
    \item Experiment 2 involves the placement of a stationary, non-living, human-sized obstacle to investigate its effects on gait features. In addition, this also helps us characterize the extent of personal space maintained by the volunteers to avoid collision in the presence of a static obstacle.
    \item Experiment 3 is performed in the presence of a human being as an obstacle (a living obstacle) to observe the changes in the behavior of the participants to the presence of a living entity instead of a non-living obstacle. The change in personal space due to the presence of another human is also characterized using the trajectories from this experiment.
\end{itemize}

The experiments are conducted utilizing a group of volunteers, including both males and females, to investigate the role of the gender of the participants on the gait features. During the experiment, each volunteer's gender, height, and shoulder length were measured and recorded for additional analysis, refer to Table \ref{table:voluteers}. All three experiments were performed sequentially for a single participant at a time. Hence, it is safe to assume that the experiments were conducted under the same surrounding conditions, thereby enabling a fair comparison of the results.

\section{Image Processing} \label{sec: ImageAnalysis}

\begin{figure}

    \centering
    \includegraphics[width= 1\linewidth]{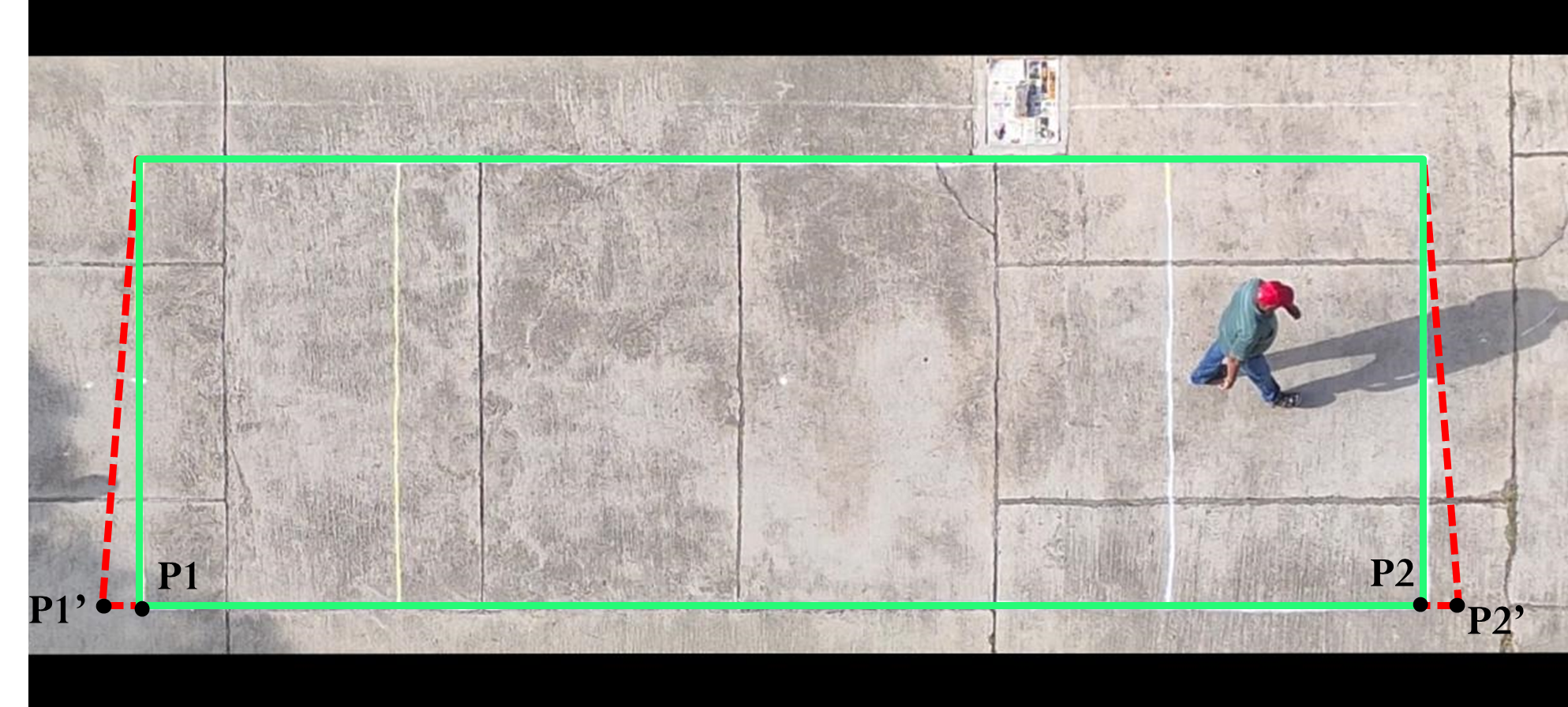}
    \caption{{\footnotesize The figure depicts the change in the rectangular corridor before (red dashed line) and after (green solid line) applying Homographic Transformation (HT). The difference between the two may be seen by considering the points P1, P1' and P2, P2'. The sum of the distance between points $P1$ $P1'$ and points $P2$ $P2'$ is about $10\%$ of the length of the corridor. Hence, the use of homographic transformation is necessary to preserve the accuracy of final trajectories.}} \label{fig:HomographicTransformation}
\end{figure}

Before the extraction of pedestrian trajectories, it is necessary to perform perspective correction on all the frames in the videos to correct distortions caused by the camera angle. Perspective correction can be obtained by using \textit{Homographic Transformation} (HT). In this transformation, the equidistance of points is preserved. The homographic transformation of one of the frames is depicted in Figure \ref{fig:HomographicTransformation}. The figure shows the experimental corridor before  (red dashed lines) and after HT (green solid lines). The difference between the two may be seen by considering the points P1, P1' and P2, P2'. The sum of the distance between points $P1$ $P1'$ and points $P2$ $P2'$ is about $10\%$ of the length of the corridor. Hence, the use of homographic transformation is necessary to preserve the accuracy of final trajectories.

\begin{figure}

    \includegraphics[width= 1\linewidth]{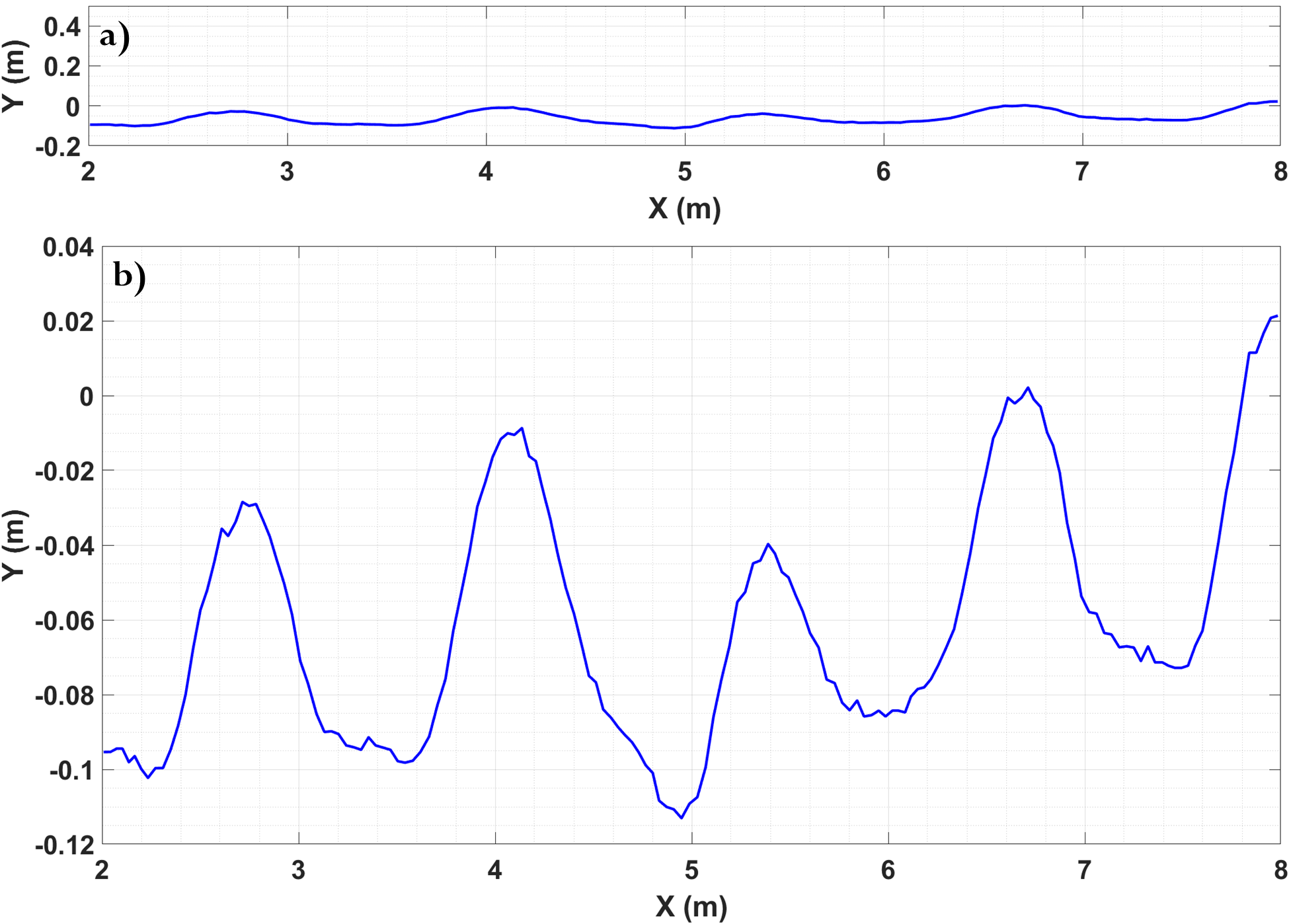}
    \caption{{\footnotesize 
    a) Sample trajectory after performing the perspective correction. b) The trajectory with an enlarged scale of the Y-axis clearly shows periodic wave-like characteristics.}} \label{fig:trajectoryExtraction2}
\end{figure}

After applying perspective correction, red caps are detected in each frame to extract the positions of volunteers. This process enables the generation of trajectories for all volunteers from the recorded video. However, the generated trajectories may contain disturbances. Therefore, post-processing is performed to remove noise and enhance accuracy.

\begin{figure}
    
    \centering
    \includegraphics[width= 1\linewidth]{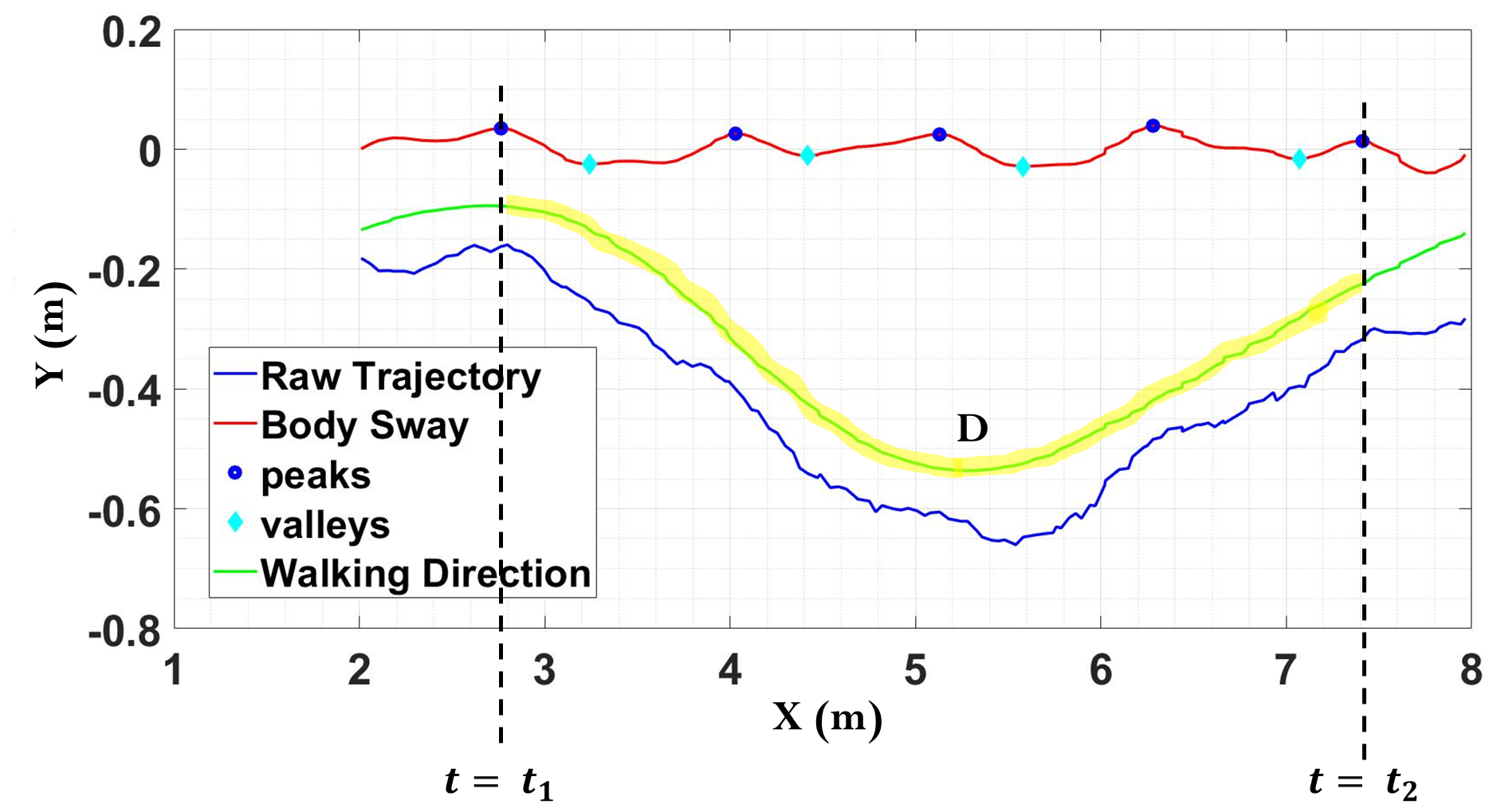}
    \caption{{\footnotesize The figure shows the graph of an original (raw) trajectory as the combination of walking direction (WD) and body sway (BS). The trajectory is extracted from experiment 2 where an obstacle is placed in the middle of the path. Blue circles represent the peaks of gait cycles, while the cyan diamonds depict the corresponding valleys in the BS trajectory. $t_2-t_1$ is the time interval between the first and last peaks. The highlighted region in the WD trajectory shows the direct distance (D). Note that WD is shifted upwards in this graph to differentiate it from the original trajectory.}} \label{fig:WDBSpeakValley}
\end{figure}

As mentioned earlier, a pedestrian's trajectory, rather than a straight line, shows periodic wave-like characteristics. In experiment 1, with the absence of any obstacle, Figure \ref{fig:trajectoryExtraction2} shows periodic wave-like characteristics in the XY plot. This periodicity is a result of the natural oscillatory motion of non-disabled human beings to balance the body mass while alternating the movement of their left and right legs to enable forward motion \cite{jia2019experimental}. According to the existing literature, pedestrian trajectories consist of two main types of information, \textit{Body Sway} (BS) and the primary \textit{Walking Direction} (WD), along with some high-frequency noise \cite{jelic2012properties, jia2019experimental, parisi2016experimental}. For instance, Figure \ref{fig:WDBSpeakValley} shows a raw trajectory extracted from experiment 2, where an obstacle is placed in the middle of the path. The graph also shows the BS trajectory, displaying lateral movements of the body and the WD trajectory, displaying the main path taken by the pedestrian without any sway (*further details provided in the figure will be discussed in the next section).

\begin{figure}

    \centering
    \includegraphics[width= 1\linewidth]{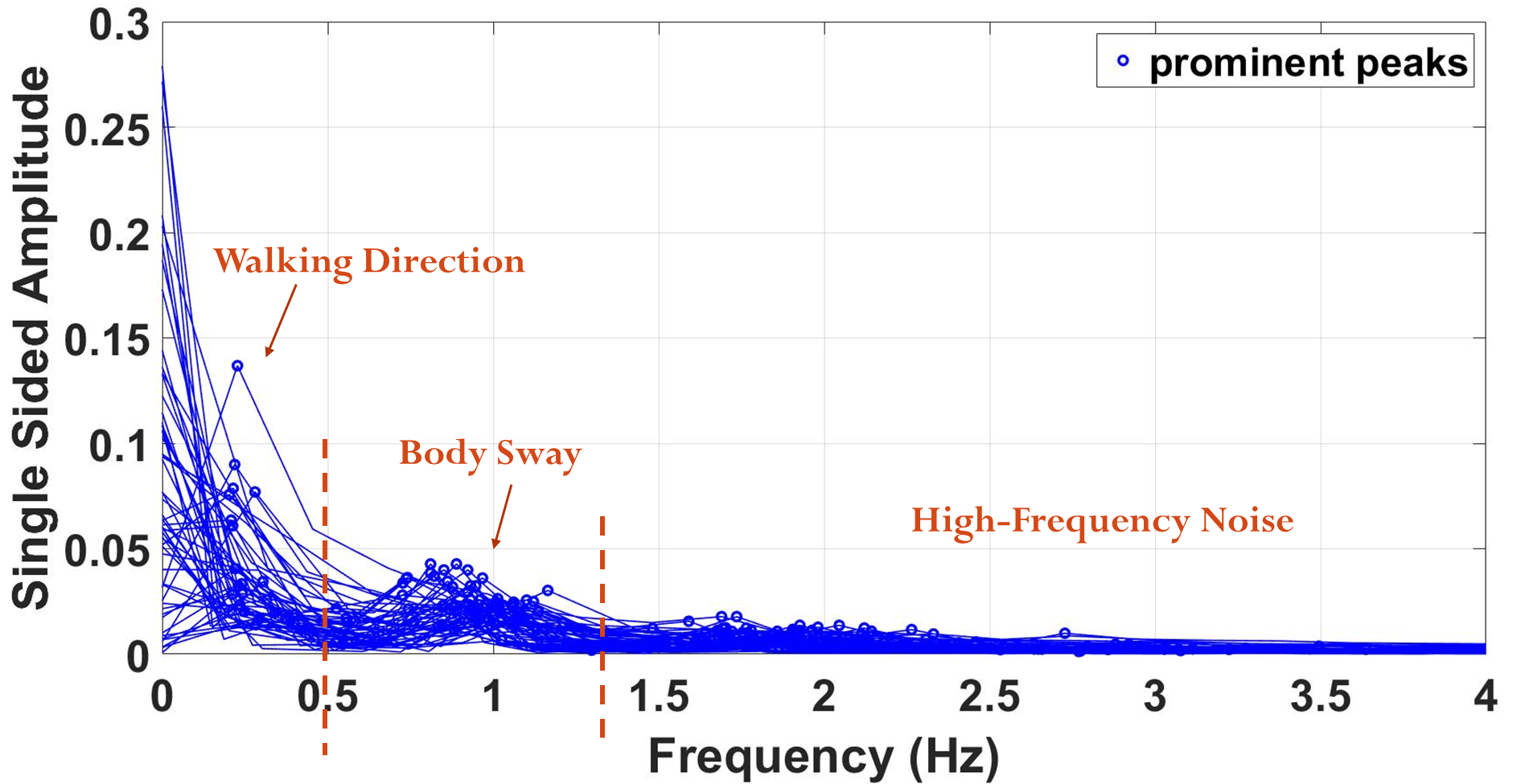}
    \caption{{\footnotesize Frequencies and corresponding amplitudes are extracted from all trajectories, with the top four prominent peaks for each trajectory highlighted by blue dots. Three regimes are visible, displaying three sets of frequencies for walking direction, body sway, and high-frequency noise, respectively.}} \label{fig:trajectoryAsCombinationOfFreq}
\end{figure}

Recent studies have established the usage of the Fourier transforms to extract these two characteristics of motion \cite{jia2019experimental, Kale2017novel, parisi2016experimental}. In Figure \ref{fig:trajectoryAsCombinationOfFreq}, we plot the frequencies and corresponding amplitudes extracted from all trajectories, highlighting the top four prominent peaks for each trajectory with blue dots. The graph is divided into three zones: the first zone represents frequencies associated with WD, the second zone shows frequencies associated with BS, and the third zone contains high-frequency noise, which is addressed through noise removal using a low-pass filter. This frequency information for WD and BS is further used for gait feature calculations.

\begin{figure*}
    \centering
    \includegraphics[width= 0.95\linewidth]{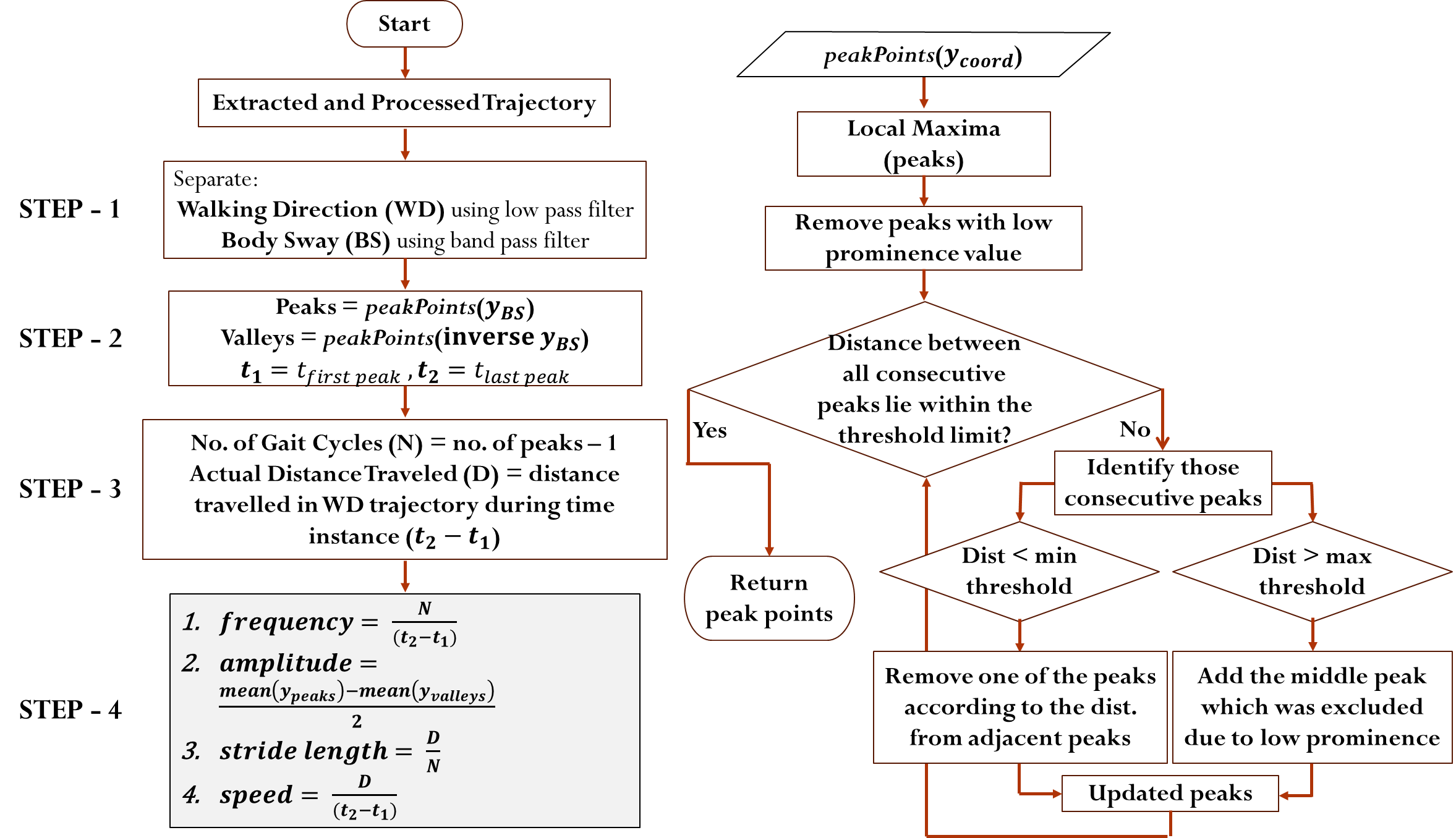}
    \caption{{\footnotesize The flow chart (left) illustrates the algorithm for automatic computation gait features from a pedestrian trajectory obtained using a single camera. This flowchart depicts the step-by-step process utilized to extract various gait features from the trajectory data of a pedestrian. The algorithm involves several stages including preprocessing, peak detection, and feature computation, aimed at accurately characterizing the individual's gait pattern. The flow chart (right) defines the 'peakPoints' function, identifying peaks and valleys within the pedestrian's trajectory. Peaks and valleys indicate the steps taken by the pedestrian. For details of the algorithm steps, refer to text.}} \label{fig:GFflowchart}
\end{figure*}

\section{Calculation of Gait features} \label{sec:gaitFeatures}

In this section, the process of calculating gait features, including body sway frequency, amplitude, stride length and speed, is explained. As mentioned earlier, the two components of a trajectory - WD and BS, can be generated with the help of a set of frequencies. BS contains information of lateral movements due to the postural adjustment and mainly contributes to the calculation of body sway frequency and amplitude. WD highlights the main path taken by the pedestrian to reach the goal, and it contributes to the calculation of stride length and speed. In order to calculate the gait features of a trajectory extracted from a video shot, an algorithm is provided in this study. A detailed flow chart summarizing the algorithm is provided in Figure \ref{fig:GFflowchart}.

The algorithm consists of the following steps:

Step 1: The separation of the walking direction (WD) and body sway (BS) from a trajectory is performed by using low pass filter and band pass filter, respectively.

Step 2: The identification of local maxima (peaks) and minima (valleys) is performed on BS trajectory using our `peakPoints' function described in Figure \ref{fig:GFflowchart} (right). The peaks and valleys are carefully selected in order to maintain an approximate distance between two consecutive peaks. The time instances at the first and the last peaks are stored as $t_1$ and $t_2$, respectively.

Step 3: Each gait cycle has a series of peak-valley-peak points as was shown in Figure \ref{fig:gaitCycle}. In this study a gait cycle is assumes to always start and end with a peak at time instances $t_1$ and $t_2$. Hence, the number of gait cycles (N) in a trajectory is one less than the number of peaks. The actual distance traveled (D) is the distance traveled in the time $t_2-t_1$ in WD trajectory. Refer to Figure \ref{fig:WDBSpeakValley} for more details.

Step 4: Frequency is calculated by dividing the number of gait cycles (N) by the time interval ($t_2-t_1$). The amplitude is calculated by subtracting the mean of the Y-coordinates at the valleys from the mean of the Y-coordinates at the peaks and dividing the value by 2. The stride length is calculated by dividing the actual distance traveled (D) by the number of gait cycles (N). Lastly, the speed is calculated by dividing the distance traveled (D) by the time interval ($t_2-t_1$).

Note that in our ‘peakPoints’ function, the threshold prominence value used is $10^{-2}$. This is selected by studying the effect of various prominence values on the calculations. The upper and lower threshold values were determined by manually estimating the stride lengths throughout the experiment which are 0.9 m and 2.1 m, respectively.

It should also be noted that the separation of body sway (BS) and walking direction (WD) is a crucial step, particularly when the direction of the trajectory is changing to avoid a collision. This is especially relevant in our scenario, including experiments 2 and 3, where obstructions are present in the middle of the path. In this case, the task of identifying gait cycles becomes very challenging due to the variable nature of the walking direction, whose example can be seen in Figure \ref{fig:WDBSpeakValley}. This is where the Fourier Spectrum becomes essential, where the initial set of frequencies provides the walking direction, which can be extracted using a low pass filter, and the next set of frequencies provides body sway, which can be extracted using a band pass filter.

\begin{figure} 
  
    \centering
    \includegraphics[width= 0.9\linewidth]{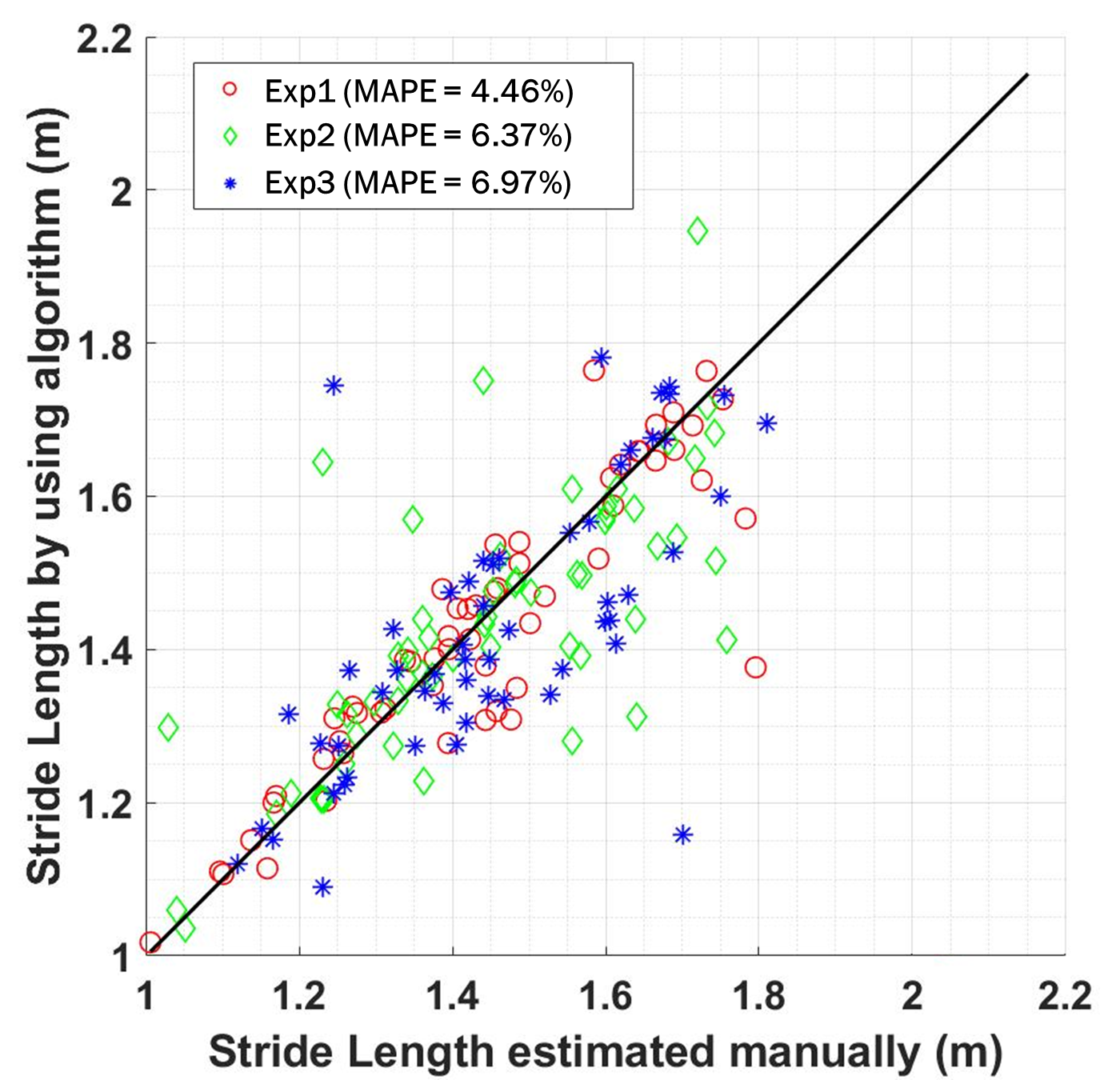}
    \caption{{\footnotesize The graph shows a comparison of stride length calculated using the automatic algorithm with the stride length calculated manually. The mean absolute percentage error (MAPE) is reported as less than 10\% for the three experiments, which shows that the algorithm exhibits a high level of consistency.}} \label{fig:videoVSalgo}
\end{figure}

\begin{figure*}
    \captionsetup{justification=justified,singlelinecheck=false} 
    \centering
    \includegraphics[width= 1\linewidth]{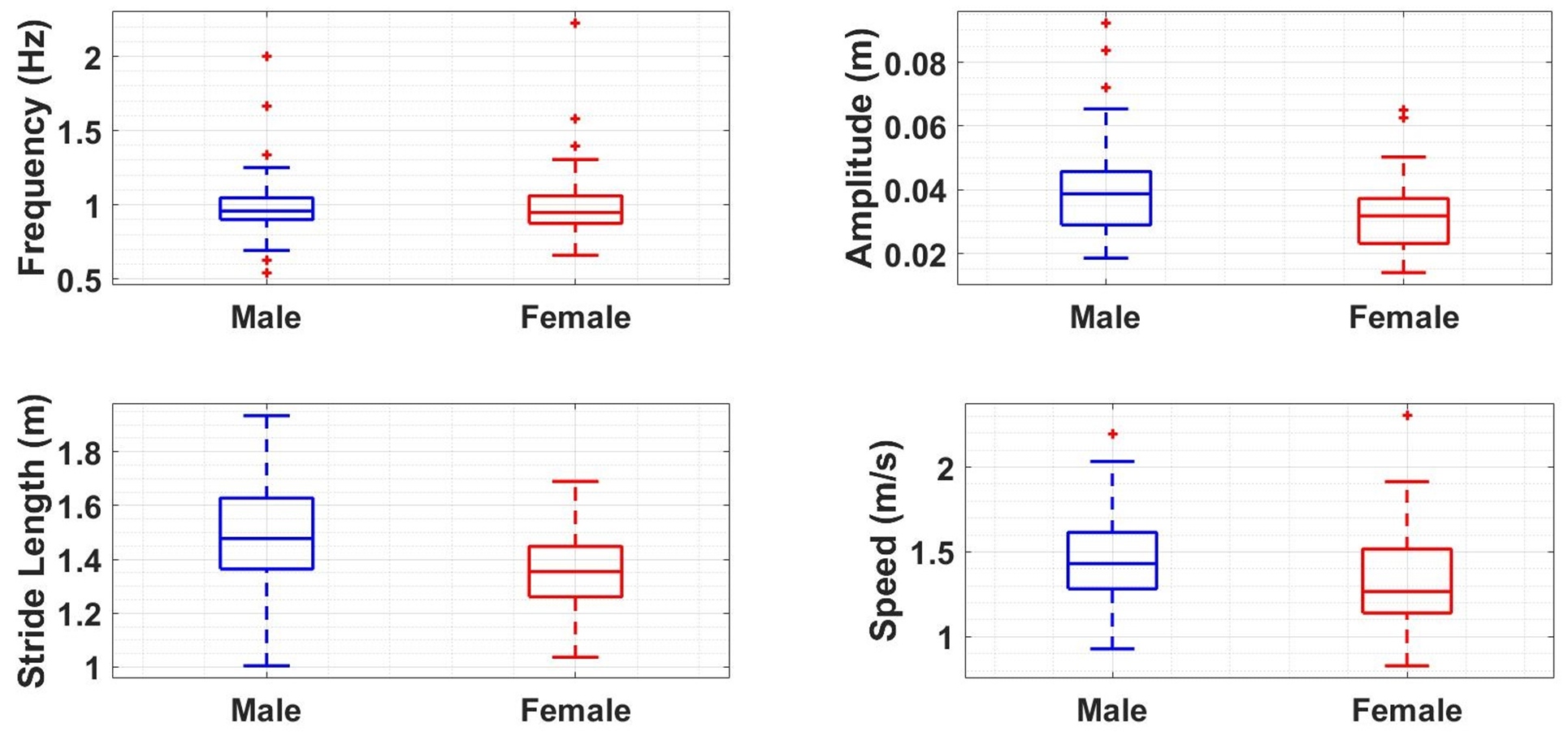}
    \caption{{\footnotesize The figure shows the box plots highlighting gait features with respect to gender from the data collected. However, to test the statistical significance of the difference between the groups, the Mann-Whitney U test is done as presented in Table \ref{table:gaitFeaturesKW}.}} \label{fig:gaitFeatues}
\end{figure*}

\begin{table*}
\centering

\def\arraystretch{2}

\scriptsize

\centering

\caption{{\footnotesize Percentage difference of the median of gait features with respect to gender, presence of an obstacle, and the type of obstacle. Percentage difference = $100 \times \frac{|med{A} - med{B}|}{\frac{med{A} + med{B}}{2}}$. The group differences are examined using the Mann-Whitney U test for the group median. In the table, *, **, and *** imply significance at the levels 0.05, 0.01, and 0.001, respectively. The results show significant differences in gait features between males and females, particularly in amplitude, stride length, and speed. However, no significant changes in gait features are observed in the presence of obstacles, whether living or non-living.}}

\begin{tabular}{|c|c|c|c|}
\hline
& $\textbf{Males (A)}$ & $\textbf{Without obstacle (A)}$ & $\textbf{Non-living obstacle (A)}$\\
& $\textbf{vs Females (B)}$ & $\textbf{vs With obstacle (B)}$ & $\textbf{vs Human obstacle (B)}$\\\hline
$\textbf{Frequency}$ & 1.06\% & 2.1\% & 0.83\%\\\hline
$\textbf{Amplitude}$ & $28.64\%^{***}$ & 20.59 & 1.15\%\\\hline
$\textbf{Stride Length}$ & $8.68\%^{**}$ & 0.46\% & 0.13\%\\\hline
$\textbf{Speed}$ & $8.14\%^*$ & 3.95\% & 2.74\%\\\hline

\end{tabular}
\label{table:gaitFeaturesKW}

\end{table*} 

In Figure \ref{fig:videoVSalgo}, the manual estimation of stride length is compared with the stride length calculated using the algorithm. The algorithm exhibits a high degree of consistency with a mean absolute percentage error (MAPE) of $4.46$\%, $6.37$\%, and $6.97\%$ for Experiments 1, 2, and 3, respectively.

The mean body sway \textbf{frequency} for trajectories with no obstructions is found to be approximately \bm{$0.99 \pm 0.16\;Hz$},  which agrees well with frequencies reported in previous studies, specifically $0.95\;Hz$ \cite{Kale2017novel}, range $0.72-1.04$ \cite{zhang2018height}, and $0.9\;Hz$ \cite{jia2019experimental}. Previous studies have shown that the amplitude of sway exhibited by an individual, walking at a normal pace, ranges from $25.5\;mm$ to $37\;mm$ \cite{wang2010body} and may reach up to $40\;mm$ \cite{jia2019experimental}. Our algorithm reports the average \textbf{amplitude} as \bm{$32.3 \pm 0.009\;mm$}. Note that defining stride length is itself a subject of ongoing research \cite{sekiya1997optimal, yang2022real, zhang2018height}. Studies have reported varying estimates of step length, which is defined as half of the stride length. Sekiya et al. \cite{sekiya1997optimal} reported step length for men of around $0.76\;m$, while for females, it is approximately $0.69\;m$. Zhang et al. \cite{zhang2018height} reported an overall step length of 0.72 m. Additionally, Yang et al. \cite{yang2022real} observed that the step length is approximately 0.67 m for indoor walking and 0.66 m for outdoor walking. Our algorithm in this study yields an average \textbf{stride length} of \bm{$1.43 \pm 0.18\;m$} (i.e., step length of $0.71 \pm 0.09\;m$) and an average \textbf{speed} of \bm{$1.41 \pm 0.24\;m/s$}.

The differences in gait features, in terms of gender, are visually presented by box plots, shown in Fig. \ref{fig:gaitFeatues}. We observe significant variability across all gait features in the box plot. However, to test the statistical significance of the difference between the groups, the Mann-Whitney U test is done as presented in Table \ref{table:gaitFeaturesKW}. The table also investigates if the gait features are also affected by the surrounding factors, such as the existence of an obstacle and the type of obstacle (non-living and human). The test indicates that there is no significant difference in frequency between males and females, while there exist significant differences in amplitude, stride length, and speed, with typically men exhibiting larger values than females. Specifically, females exhibit 8.68\% smaller stride lengths, 8.14\% slower speeds, and 28.64\% smaller amplitudes in comparison to males. Further, we observed that individuals exhibit no behavioral change in response to obstacles, regardless of their nature.

Many studies have examined gender recognition using various gait features such as hip, knee, and ankle movements \cite{mckean2007gender}, as well as head, arm, trunk, and thigh movements \cite{li2008gait}, or full-body analysis through image processing \cite{yu2009study, makihara2011gait}. This raises the question of whether differences in gait features are primarily due to gender or body parameters. To investigate this, a Mann-Whitney U test was conducted with the hypothesis that three gait features—amplitude, stride length, and speed—would continue to show differences between genders even when normalized by body parameters such as shoulder length and height. In Table \ref{table:gaitFeatures_normalised}, the amplitude is normalized by shoulder length, while stride length and speed are normalized by height. The test results indicate that body parameters act as confounding variables, primarily influencing differences in gait features rather than gender itself.

\begin{table}
\centering

\def\arraystretch{1.5}

\scriptsize

\centering

\caption{{\footnotesize Mann-Whitney U test is conducted with the hypothesis that the gait features would continue to show differences between genders even when normalized by body parameters. The amplitude is normalized by shoulder length, while stride length and speed are normalized by height. The test indicates that body parameters act as confounding variables, primarily influencing differences in gait features rather than gender itself. Since no significant difference was found in frequency between genders, no further tests were conducted for this feature.}}

\begin{tabular}{|c|c||c|c|}
\hline
& $\textbf{Males (A) vs}$ & & $\textbf{Males (A) vs}$\\
& $\textbf{Females (B)}$ & & $\textbf{Females (B)}$\\\hline
\rule{0pt}{20pt} \raisebox{+0.75\height} {$\textbf{Amplitude}$} & \raisebox{+0.75\height}{$28.64\%^{***}$} & \raisebox{+0.5\height} {$\mathbf{\dfrac{Amplitude}{Shoulder Length}}$} & \raisebox{+0.75\height}{3.6\%}\\\hline
\rule{0pt}{20pt} \raisebox{+0.75\height} {$\textbf{Stride Length}$} & \raisebox{+0.75\height}{$8.68\%^{**}$} & \raisebox{+0.5\height} {$\mathbf{\dfrac{Stride Length}{Height}}$} & \raisebox{+0.75\height}{2.65\%}\\\hline
\rule{0pt}{20pt} \raisebox{+0.75\height} {$\textbf{Speed}$} & \raisebox{+0.75\height}{$8.14\%^{*}$} & \raisebox{+0.5\height} {$\mathbf{\dfrac{Speed}{Height}}$} & \raisebox{+0.75\height}{2.73\%}\\\hline

\end{tabular}
\label{table:gaitFeatures_normalised}

\end{table} 

\section{Gait Feature Relationship}

We used multiple linear regression to determine if the predictor variables, namely, speed, gender, shoulder length, and height of pedestrians, significantly predicted the response variables: body sway frequency, amplitude, and stride length. We performed stepwise regression with a pEnter (p-value to enter), and pRemove threshold of 0.05 and 0.10, respectively. The stepwise regression involved considering an intercept term, linear and squared terms for each predictor, and interactions between pairs of distinct predictors while constructing the regression model. 

The predictor variables are speed (\textit{v}, unit: m/s), gender (\textit{g}, unit: M/F), shoulder length (\textit{l}, unit: m), and height of pedestrian (\textit{h}, unit: m). The predictor variable, `gender', is a categorical variable, and the final selected model for each dependent variable is given below.

\begin{align}
    Frequency\;(Hz) =&\;2.39 -2.09\;v - 0.8\;l + 0.93\;v^2\label{eq:freq}\\
    Amplitude\;(m) =& -0.01\;v + 0.09\;l\label{eq:amp}\\
    Stride\;Length\;(m) =& -2.01 + 3.5\;v + 0.49\;h - 1.11\;v^2\label{eq:SL}
\end{align}

As expected, gender is not significant in predicting any of the response variables. 

\begin{figure*}
    \captionsetup{justification=justified,singlelinecheck=false} 
    \centering
    \includegraphics[width= 0.8\linewidth]{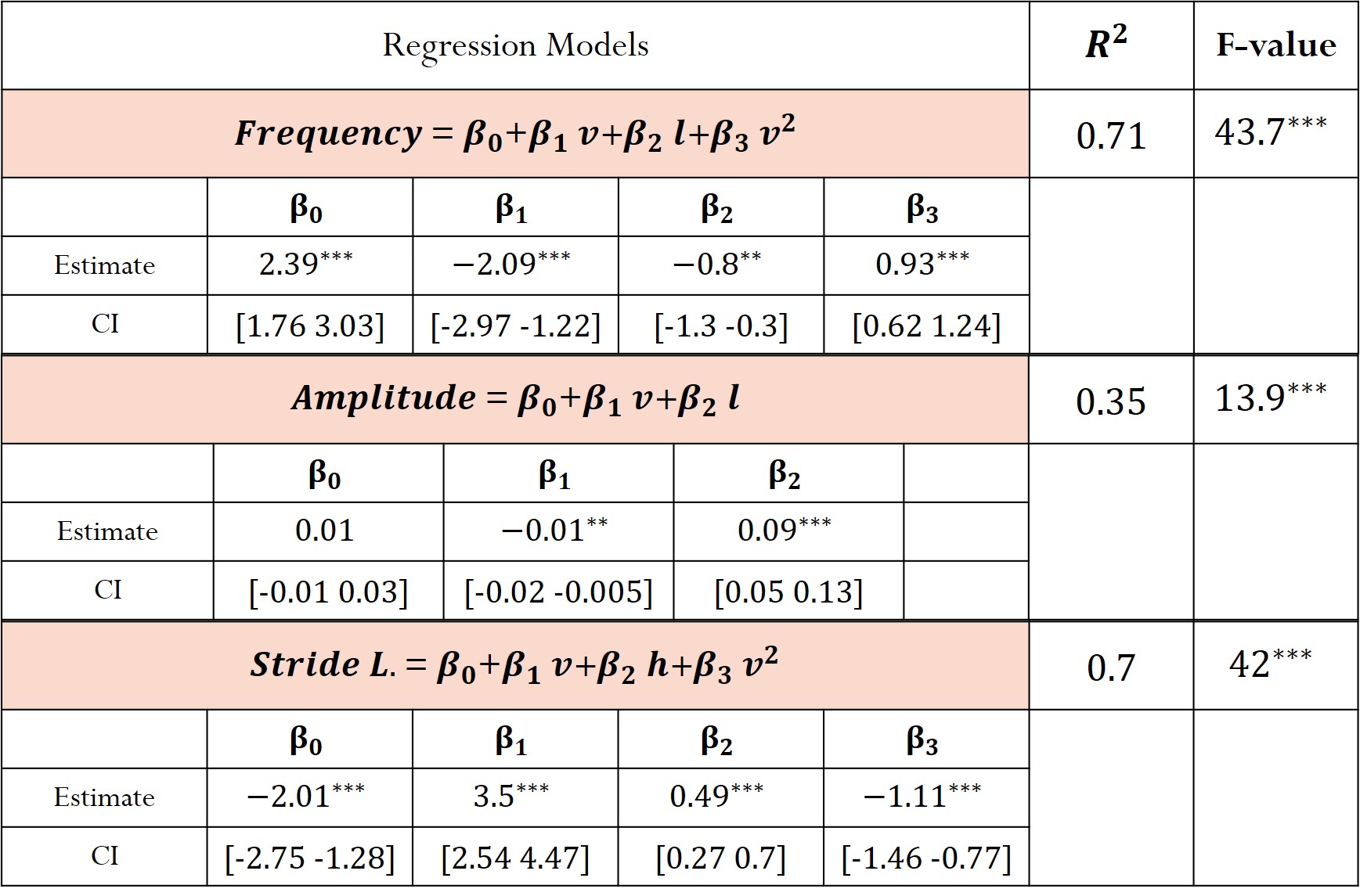}
    \caption{{\footnotesize The multiple linear regression results for frequency, amplitude, and stride length. The predictor variables are speed (\textit{v}, m/s), gender (\textit{g}, M/F), shoulder length (\textit{l}, m), and height of pedestrian (\textit{h}, m)}} \label{fig:gaitRelationAnalysis}
\end{figure*}

The regression models above are statistically significant using F-test. The $R^2$ value for BS frequency is 0.71 (eq. \ref{eq:freq}), whereas the corresponding F-statistic is 43.7. Similarly, the amplitude has $R^2$ value of 0.34 (eq. \ref{eq:amp}), and the corresponding F-statistic is 13.9. The stride length has $R^2$ value of 0.7 (eq. \ref{eq:SL}), and the corresponding F-statistic is 42. In Figure \ref{fig:gaitRelationAnalysis}, the results indicate a significant quadratic relationship of speed with frequency ($\beta = 0.93^{***},\;95\%\;CI\;[0.62,\;1.24]$), and stride length ($\beta = -1.11^{***},\;95\%\;CI\;[-1.46,\;-0.77]$). The outcomes also indicated a significant linear relationship between speed and the predicted amplitude ($\beta = -0.01^{**},\;95\%\;CI\;[-0.02,\;-0.005]$). A similar relationship has been witnessed by Jia et al. \cite{jia2019experimental}, but the author did not consider non-linear features, and no rigorous statistical analysis was conducted. Our results indicate that there is a significant linear relationship between shoulder length and anticipated frequency ($\beta = -0.8^{**},\;95\%\;CI\;[-1.3,\;-0.3]$), as well as between shoulder length and amplitude ($\beta = -0.09^{***},\;95\%\;CI\;[0.05,\;0.13]$). We also found a significant relation between height and stride length ($\beta = 0.49^{***},\;95\%\;CI\;[0.27,\;0.7]$).

The model predictions and the raw data are also plotted in the supporting Figure S1, where graphs on the left depict an isometric view, while those on the right display the corresponding front view, displaying a near-perfect agreement. For an additional robustness check, we also performed the robust regression, and the results are reported in the supporting Figure S2.

\section{Personal Gap}

The effect of obstacles on the personal space kept by pedestrians in a crowd has been the subject of many studies \cite{yanagisawa2009introduction, yanagisawa2010study, jia2019experimental, parisi2016experimental}. Jia \cite{jia2019experimental} conducted experiments to observe the impact of a non-living obstacle placed in the path of a pedestrian on the walking direction and personal space and found some changes in the features. Several other studies have regarded a standing pedestrian as an obstacle and have documented the critical headway, which refers to the gap between a pedestrian and the obstruction when the pedestrian intends to initiate avoidance maneuvers \cite{lv2013two, moussaid2009experimental, parisi2016experimental, shan2014critical}. According to Lv \cite{lv2013two} and Moussaid \cite{moussaid2009experimental}, the critical headway (personal gap) values were observed to be concentrated within the ranges of $0.9\;m$ to $2.0\;m$ and $1.5\;m$ to $2.5\;m$, respectively. Parisi \cite{parisi2016experimental} determined that the minimal distance necessary to prevent collision was determined to be no more than $1\;m$. Previous studies have demonstrated that an individual requires a minimum space of approximately $2\;m^2$ to navigate around a stationary person safely \cite{shan2014critical}. Furthermore, while encountering a pedestrian moving in the opposite direction, it is recommended to have a space of approximately $2.64\;m^2$ to ensure avoidance. 

\begin{figure} [h]
    
    \centering
    
    \subfigure[]
    {
        \includegraphics[width= 1\linewidth]{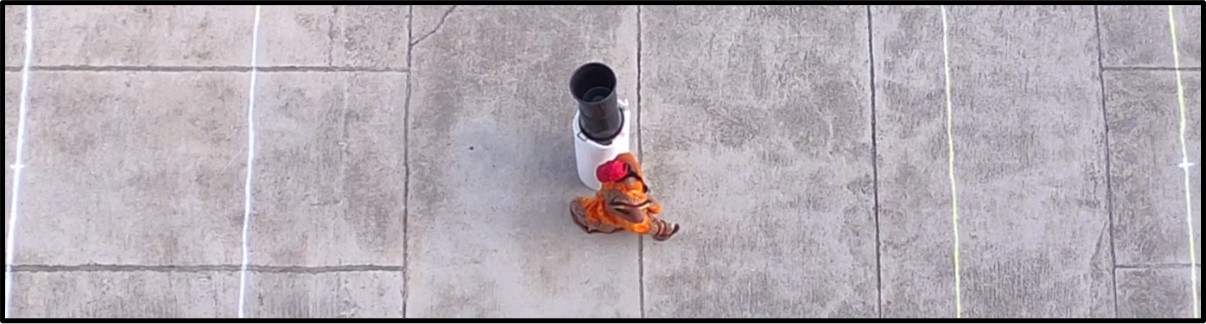}
        \label{fig:stationaryObst}
    }

    \subfigure[]
    {
        \includegraphics[width= 1\linewidth]{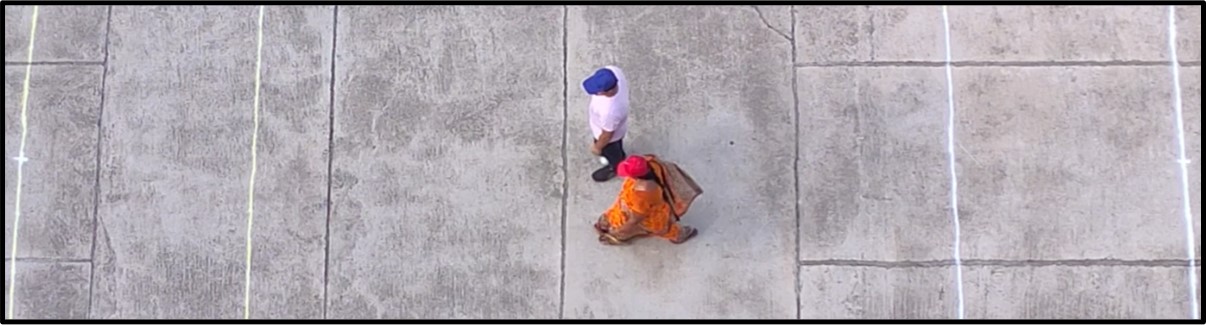}
        \label{fig:standingPerson}
    }

    \caption{{\footnotesize Two different experimental scenarios with an obstacle: a) a stationary, non-living, human-sized cylindrical obstacle is present in the path, b) a human being (male) is standing as an obstacle in the path.}}
    \label{fig:E2E3}
\end{figure}

\begin{figure}
    
    \centering
    \includegraphics[width= 1\linewidth]{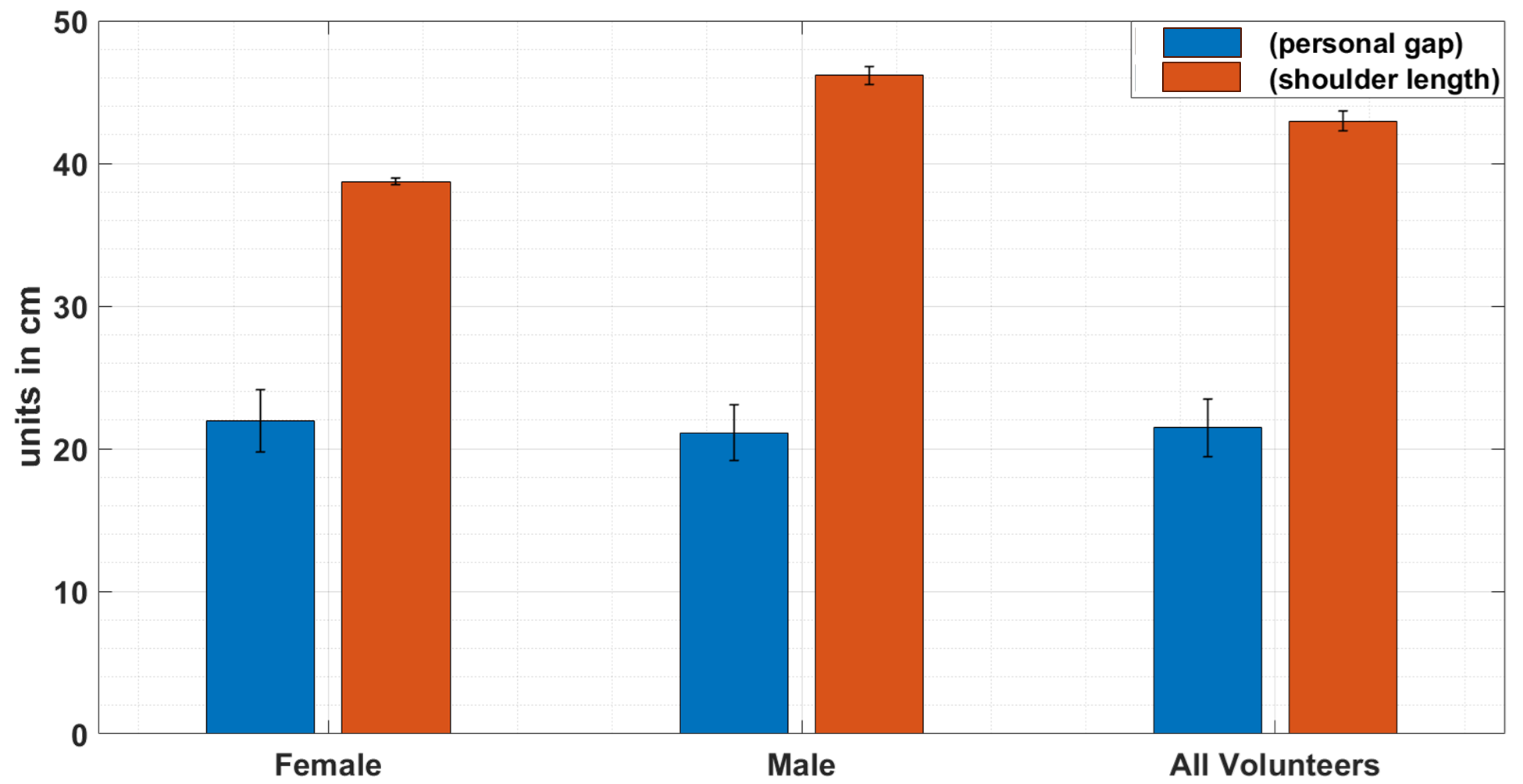}
    \caption{{\footnotesize The mean personal space and shoulder length for volunteers in experiments.}} \label{fig:personalSpace}
\end{figure}

In this study, we have analyzed the personal gap maintained by the volunteers for two types of obstructions. The first is a stationary, non-living-human-sized cylinder (Fig. \ref{fig:stationaryObst}, whereas the second involves a standing human being, as seen in Fig. \ref{fig:standingPerson}. Our results show that, on an average, a pedestrian tends to maintain a personal gap of about half the size of his/her own physical dimensions, namely their shoulder width. The ratio between the average personal gap and half of the average shoulder length is found to be $0.99$, and the Wilcoxon Signed-Rank test for the difference in these values has $p-value$ of $0.64$. While going gender specific, the ratios of the averages are $1.13$ and $0.91$ for females and males, respectively (see Figure \ref{fig:personalSpace}).

\begin{table*}
\centering

\def\arraystretch{1.3}

\scriptsize
\caption{{\footnotesize The table includes the percentage difference of the median of the personal gap with respect to gender in experiments 2 and 3. Percentage difference is calculated as $100 \times \frac{|med{A} - med{B}|}{\frac{med{A} + med{B}}{2}}$. The group differences are examined using the Mann-Whitney U test for the group median. In the table, *, **, and *** imply significance at the levels 0.05, 0.01, and 0.001, respectively. Observation is given for each test, and the results conclude that pedestrians treat stationary human beings the same as stationary non-living obstacles, given that the gender is the same to match the comfort level.}}
\label{table:personalSpace}

\begin{tabular}{|c|c|l|l|}
  \hline
   & $\textbf{\%\;Difference}$ & \textbf{Observations} & \textbf{Conclusion}   \\
  \hline
  $\textbf{E2M, E2F}$ & 3.81\%  & Males and females maintained a similar personal&  \\
            &    & gap from the stationary obstacle & \\
    & & & \\
    
  $\textbf{E2M, E3M}$ & 1.26\%  & Males maintained the similar gap for both stationary & \\
        &    & obstacle and standing human being (male) & \begin{small}Pedestrians treat stationary human\end{small}\\

& & & \begin{small}being and stationary obstacle the same \end{small}.\\
        
  $\textbf{E2F, E3F}$ & $38.16\%^{*}$  & Females showed significant change in their personal & \begin{small}given that the gender is same to match\end{small}\\
            &    & gap in the presence of an opposite gender. & \begin{small}the comfort level.\end{small}\\
            
  & &  & \\
  $\textbf{E3F, E3M}$ & $43.03\%^{**}$ & Males maintained their personal gap, while the & \\
        & & females increased theirs in the presence of male & \\
        & & volunteer as an obstacle. & \\
  \hline
\end{tabular}

\end{table*}

The personal gaps maintained in both experiments 2 and 3 are also analyzed statistically with respect to gender. For the testing, four groups are formed - experiment 2 males (E2M), experiment 2 females (E2F), experiment 3 males (E3M), and experiment 3 females (E3F). In Table \ref{table:personalSpace}, with the help of percentage difference, E2M and E2F are not significantly different, i.e., the personal gap maintained by males and females from the stationary obstacle is similar. E2M and E3M are not significantly different, i.e., males maintained a similar gap for both stationary obstacle and standing human being (male). E2F and E3F show significant difference, i.e., females showed a change in their personal gap in the presence of an opposite gender. E3F and E3M show significant difference, i.e., males maintained their personal gap, while the females increased theirs in the presence of male volunteer as an obstacle. The whole exercise gives an important conclusion that pedestrians treat stationary human being and stationary obstacle the same given that the gender is same to match the comfort level. Therefore, a non-living obstacle can be used as a proxy for a standing living person for future experiments. These personal space measurements and observations are expected to be important inputs to pedestrian dynamics simulations \cite{kramer2021social}.

\section{Choice of Direction}

Despite the fact that our experimental scenario is symmetric, pedestrians may exhibit a preference for either turning left or right before passing the obstacle. The study by Jia et al.\cite{jia2019experimental} also examined the decision-making process involved in selecting the left or right directions. Our results reveal that a majority, over 70 percent of the participants, exhibit a preference for turning left, regardless of whether the obstacle is a human or a non-living one (see Table \ref{table:leftRight}). Consequently, this information can be used to set the bias to turn left in any pedestrian dynamics model for the Indian scenario. Note this bias may originate from the left-hand driving rules prevalent in India. We believe it may hold for other countries with similar left-driving rules. Jia \cite{jia2019experimental} has reported similar biases in the Japanese context.

\begin{table}
\centering

\def\arraystretch{2}

\scriptsize

\caption{{\footnotesize The left and right choices made by the volunteers while passing the obstacle in experiment 2 and experiment 3.}}
\label{table:leftRight}

\begin{tabular}{|c|c|c|}
\hline
 & $\textbf{Left}$ & $\textbf{Right}$\\\hline
$\textbf{E2}$ & 54 & 20\\\hline
$\textbf{E3}$ & 50 & 24\\\hline

\end{tabular}

\end{table} 

\section{Summary and Conclusions}

The study introduces an algorithm (Figure \ref{fig:GFflowchart}) for the calculation of gait features of a trajectory extracted from a short video clip. To analyze the change in gait features with respect to gender as well as the presence of obstructions, three different experiments are performed in which the volunteers are walking towards the destination: a) without any obstruction, b) with a stationary non-living obstacle present in the middle of the path, and c) with a human being standing in the middle of the path. The key conclusions from the analysis are as follows.

\begin{itemize}
    \item The gait features determined in this study are - frequency ($0.9896 \pm 0.1645\;Hz$), amplitude ($32.3 \pm 0.009\;mm$), stride length ($1.43 \pm 0.18\; m$), and speed ($1.41 \pm 0.24\;m/s$).
    \item The presence or the nature of obstruction does not seem to have an impact on the gait features (Table \ref{table:gaitFeaturesKW}). This finding suggests that a pedestrian anticipates the obstruction present in his path and adjusts the walking direction, therefore maintaining the flow of movement.
    \item There are notable differences between males and females in terms of stride length, speed, and amplitude. Specifically, females exhibit 8.68\% smaller stride lengths, 8.14\% slower speeds, and 28.64\% smaller amplitudes in comparison to males, with no significant difference in frequency. However, according to further investigation, our study reveals that the body parameters are the main variables that dominate gait features rather than gender (Table \ref{table:gaitFeatures_normalised}).
\end{itemize}

In this study, we have also analyzed the personal gap maintained by the volunteers and the directional bias in the presence of an obstruction. The key conclusions from the analysis are as follows.

\begin{itemize}
    \item The study reveals that pedestrians tend to maintain a personal gap of about half the length of their shoulders (Figure \ref{fig:personalSpace}).
    \item Pedestrians treat stationary human being and stationary obstacle the same, given that the gender is the same to match the comfort level (Table \ref{table:personalSpace}). Therefore, a non-living obstacle can be used as a proxy for a standing living person for future experiments.
    \item Our results also provide a measure of the bias (about 70\%) to turn left while walking towards an obstacle (Table \ref{table:leftRight}). The left bias is most likely due to the left-handed driving rules followed in India.
\end{itemize}

\section{Acknowledgement}

The authors would like to thank several faculties and students of IIT Kanpur, who participated in the study as volunteers. We would also like to express our sincere gratitude to the security section and personnel of IIT Kanpur for helping us in all our experiments and the IIT Kanpur administration for facilitating the same. We also acknowledge the foundations laid by previous dual degree students of IIT Kanpur - Ishan Prashant, Amullya Kale and Satyendra Pandey, towards this study. The authors acknowledge the support by the CRG project (Sanction number$<$IME/SERB/2023137$>$) from SERB, DST (India) for this study.\\\\
$^*$Email IDs to the corresponding authors are indrasd@iitk.ac.in, anuragt@iitk.ac.in, and sprawesh@iitk.ac.in.


\bibliographystyle{ieeetr}
\bibliography{references}

\begin{thebibliography}{10}

\bibitem{racic2009experimental}
V.~Racic, A.~Pavic, and J.~Brownjohn, ``Experimental identification and analytical modelling of human walking forces: Literature review,'' {\em Journal of Sound and Vibration}, vol.~326, no.~1-2, pp.~1--49, 2009.

\bibitem{perry2024gait}
J.~Perry and J.~M. Burnfield, ``Gait cycle,'' in {\em Gait Analysis}, pp.~3--6, CRC Press, 2024.

\bibitem{yoo2005gender}
J.-H. Yoo, D.~Hwang, and M.~S. Nixon, ``Gender classification in human gait using support vector machine,'' in {\em Advanced Concepts for Intelligent Vision Systems: 7th International Conference, ACIVS 2005, Antwerp, Belgium, September 20-23, 2005. Proceedings 7}, pp.~138--145, Springer, 2005.

\bibitem{jia2019experimental}
X.~Jia, C.~Feliciani, D.~Yanagisawa, and K.~Nishinari, ``Experimental study on the evading behavior of individual pedestrians when confronting with an obstacle in a corridor,'' {\em Physica A: Statistical Mechanics and its Applications}, vol.~531, p.~121735, 2019.

\bibitem{zhang2018height}
Y.~Zhang, Y.~Li, C.~Peng, D.~Mou, M.~Li, and W.~Wang, ``The height-adaptive parameterized step length measurement method and experiment based on motion parameters,'' {\em Sensors}, vol.~18, no.~4, p.~1039, 2018.

\bibitem{wang2010body}
F.~Wang, M.~Skubic, C.~Abbott, and J.~M. Keller, ``Body sway measurement for fall risk assessment using inexpensive webcams,'' in {\em 2010 Annual International Conference of the IEEE Engineering in Medicine and Biology}, pp.~2225--2229, IEEE, 2010.

\bibitem{sekiya1997optimal}
N.~Sekiya, H.~Nagasaki, H.~Ito, and T.~Furuna, ``Optimal walking in terms of variability in step length,'' {\em Journal of Orthopaedic \& Sports Physical Therapy}, vol.~26, no.~5, pp.~266--272, 1997.

\bibitem{yang2022real}
Z.~Yang, L.~C. Tran, F.~Safaei, A.~T. Le, and A.~Taparugssanagorn, ``Real-time step length estimation in indoor and outdoor scenarios,'' {\em Sensors}, vol.~22, no.~21, p.~8472, 2022.

\bibitem{thevenon2015collection}
A.~Thevenon, F.~Gabrielli, J.~Lepvrier, A.~Faupin, E.~Allart, V.~Tiffreau, and V.~Wieczorek, ``Collection of normative data for spatial and temporal gait parameters in a sample of french children aged between 6 and 12,'' {\em Annals of physical and rehabilitation medicine}, vol.~58, no.~3, pp.~139--144, 2015.

\bibitem{cai2017single}
X.~Cai, G.~Han, X.~Song, and J.~Wang, ``Single-camera-based method for step length symmetry measurement in unconstrained elderly home monitoring,'' {\em IEEE Transactions on Biomedical Engineering}, vol.~64, no.~11, pp.~2618--2627, 2017.

\bibitem{chatterjee2020analysis}
S.~Chatterjee, ``Analysis of the human gait rhythm in neurodegenerative disease: A multifractal approach using multifractal detrended cross correlation analysis,'' {\em Physica A: Statistical Mechanics and its Applications}, vol.~540, p.~123154, 2020.

\bibitem{bartsch2007fluctuation}
R.~Bartsch, M.~Plotnik, J.~W. Kantelhardt, S.~Havlin, N.~Giladi, and J.~M. Hausdorff, ``Fluctuation and synchronization of gait intervals and gait force profiles distinguish stages of parkinson's disease,'' {\em Physica A: Statistical Mechanics and its Applications}, vol.~383, no.~2, pp.~455--465, 2007.

\bibitem{abdulrehem2009low}
M.~M. Abdulrehem and E.~Ott, ``Low dimensional description of pedestrian-induced oscillation of the millennium bridge,'' {\em Chaos: An Interdisciplinary Journal of Nonlinear Science}, vol.~19, no.~1, 2009.

\bibitem{belykh2016bistable}
I.~V. Belykh, R.~Jeter, and V.~N. Belykh, ``Bistable gaits and wobbling induced by pedestrian-bridge interactions,'' {\em Chaos: An Interdisciplinary Journal of Nonlinear Science}, vol.~26, no.~11, 2016.

\bibitem{renaudin2012step}
V.~Renaudin, M.~Susi, and G.~Lachapelle, ``Step length estimation using handheld inertial sensors,'' {\em Sensors}, vol.~12, no.~7, pp.~8507--8525, 2012.

\bibitem{alvarez2006comparison}
D.~Alvarez, R.~C. Gonz{\'a}lez, A.~L{\'o}pez, and J.~C. Alvarez, ``Comparison of step length estimators from weareable accelerometer devices,'' in {\em 2006 International Conference of the IEEE Engineering in Medicine and Biology Society}, pp.~5964--5967, IEEE, 2006.

\bibitem{h2015lateral}
G.~H~Goldsztein, ``Lateral oscillations of the center of mass of bipeds as they walk. inverted pendulum model with two degrees of freedom,'' {\em AIP Advances}, vol.~5, no.~10, 2015.

\bibitem{weinberg2002using}
H.~Weinberg, ``Using the adxl202 in pedometer and personal navigation applications,'' {\em Analog Devices AN-602 application note}, vol.~2, no.~2, pp.~1--6, 2002.

\bibitem{shin2011adaptive}
S.~H. Shin and C.~G. Park, ``Adaptive step length estimation algorithm using optimal parameters and movement status awareness,'' {\em Medical engineering \& physics}, vol.~33, no.~9, pp.~1064--1071, 2011.

\bibitem{yao2020robust}
Y.~Yao, L.~Pan, W.~Fen, X.~Xu, X.~Liang, and X.~Xu, ``A robust step detection and stride length estimation for pedestrian dead reckoning using a smartphone,'' {\em IEEE Sensors Journal}, vol.~20, no.~17, pp.~9685--9697, 2020.

\bibitem{sadhukhan2023irt}
P.~SADHUKHAN, S.~Mazumder, S.~Paiva, C.~Chowdhury, P.~Das, K.~Dahal, and X.~Wang, ``Irt-sd-sle: An improved real-time step detection and step length estimation using smartphone accelerometer,'' 2023.

\bibitem{wang2018linking}
J.~Wang, M.~Boltes, A.~Seyfried, J.~Zhang, V.~Ziemer, and W.~Weng, ``Linking pedestrian flow characteristics with stepping locomotion,'' {\em Physica A: Statistical Mechanics and its Applications}, vol.~500, pp.~106--120, 2018.

\bibitem{severiukhina2017study}
O.~Severiukhina, D.~Voloshin, M.~H. Lees, and V.~Karbovskii, ``The study of the influence of obstacles on crowd dynamics,'' {\em Procedia Computer Science}, vol.~108, pp.~215--224, 2017.

\bibitem{karbovskii2019impact}
V.~Karbovskii, O.~Severiukhina, I.~Derevitskii, D.~Voloshin, A.~Presbitero, and M.~Lees, ``The impact of different obstacles on crowd dynamics,'' {\em Journal of Computational Science}, vol.~36, p.~100893, 2019.

\bibitem{chen2024experimental}
J.~Chen, Q.~Luo, Q.~Wang, J.~T. Lo, and J.~Ma, ``Experimental study on individual and crowd movement features around obstacles with different shape and size,'' {\em Physica A: Statistical Mechanics and its Applications}, vol.~645, p.~129797, 2024.

\bibitem{shiwakoti2019review}
N.~Shiwakoti, X.~Shi, and Z.~Ye, ``A review on the performance of an obstacle near an exit on pedestrian crowd evacuation,'' {\em Safety science}, vol.~113, pp.~54--67, 2019.

\bibitem{lian2015experimental}
L.~Lian, X.~Mai, W.~Song, Y.~K.~K. Richard, X.~Wei, and J.~Ma, ``An experimental study on four-directional intersecting pedestrian flows,'' {\em Journal of Statistical Mechanics: Theory and Experiment}, vol.~2015, no.~8, p.~P08024, 2015.

\bibitem{yanagisawa2009introduction}
D.~Yanagisawa, A.~Kimura, A.~Tomoeda, R.~Nishi, Y.~Suma, K.~Ohtsuka, and K.~Nishinari, ``Introduction of frictional and turning function for pedestrian outflow with an obstacle,'' {\em Physical Review E}, vol.~80, no.~3, p.~036110, 2009.

\bibitem{yanagisawa2010study}
D.~Yanagisawa, R.~Nishi, A.~Tomoeda, K.~Ohtsuka, A.~Kimura, Y.~Suma, and K.~Nishinari, ``Study on efficiency of evacuation with an obstacle on hexagonal cell space,'' {\em SICE Journal of Control, Measurement, and System Integration}, vol.~3, no.~6, pp.~395--401, 2010.

\bibitem{subaih2020experimental}
R.~Subaih, M.~Maree, M.~Chraibi, S.~Awad, and T.~Zanoon, ``Experimental investigation on the alleged gender-differences in pedestrian dynamics: A study reveals no gender differences in pedestrian movement behavior,'' {\em IEEE access}, vol.~8, pp.~33748--33757, 2020.

\bibitem{zhang2016homogeneity}
J.~Zhang, S.~Cao, D.~Salden, and J.~Ma, ``Homogeneity and activeness of crowd on aged pedestrian dynamics,'' {\em Procedia computer science}, vol.~83, pp.~361--368, 2016.

\bibitem{cao2018stepping}
S.~Cao, J.~Zhang, W.~Song, R.~Zhang, {\em et~al.}, ``The stepping behavior analysis of pedestrians from different age groups via a single-file experiment,'' {\em Journal of Statistical Mechanics: Theory and Experiment}, vol.~2018, no.~3, p.~033402, 2018.

\bibitem{aghabayk2021investigation}
K.~Aghabayk, N.~Parishad, and N.~Shiwakoti, ``Investigation on the impact of walkways slope and pedestrians physical characteristics on pedestrians normal walking and jogging speeds,'' {\em Safety Science}, vol.~133, p.~105012, 2021.

\bibitem{seyfried2005fundamental}
A.~Seyfried, B.~Steffen, W.~Klingsch, and M.~Boltes, ``The fundamental diagram of pedestrian movement revisited,'' {\em Journal of Statistical Mechanics: Theory and Experiment}, vol.~2005, no.~10, p.~P10002, 2005.

\bibitem{burstedde2001simulation}
C.~Burstedde, K.~Klauck, A.~Schadschneider, and J.~Zittartz, ``Simulation of pedestrian dynamics using a two-dimensional cellular automaton,'' {\em Physica A: Statistical Mechanics and its Applications}, vol.~295, no.~3-4, pp.~507--525, 2001.

\bibitem{helbing1995social}
D.~Helbing and P.~Molnar, ``Social force model for pedestrian dynamics,'' {\em Physical review E}, vol.~51, no.~5, p.~4282, 1995.

\bibitem{hu2023anticipation}
X.~Hu, T.~Chen, and Y.~Song, ``Anticipation dynamics of pedestrians based on the elliptical social force model,'' {\em Chaos: An Interdisciplinary Journal of Nonlinear Science}, vol.~33, no.~7, 2023.

\bibitem{fiorini1998motion}
P.~Fiorini and Z.~Shiller, ``Motion planning in dynamic environments using velocity obstacles,'' {\em The international journal of robotics research}, vol.~17, no.~7, pp.~760--772, 1998.

\bibitem{hu2022effects}
X.~Hu, T.~Chen, K.~Deng, and G.~Wang, ``Effects of the direction and speed strategies on pedestrian dynamics,'' {\em Chaos: An Interdisciplinary Journal of Nonlinear Science}, vol.~32, no.~6, 2022.

\bibitem{jelic2012properties}
A.~Jeli{\'c}, C.~Appert-Rolland, S.~Lemercier, and J.~Pettr{\'e}, ``Properties of pedestrians walking in line: Fundamental diagrams,'' {\em Physical review E}, vol.~85, no.~3, p.~036111, 2012.

\bibitem{parisi2016experimental}
D.~R. Parisi, P.~A. Negri, and L.~Bruno, ``Experimental characterization of collision avoidance in pedestrian dynamics,'' {\em Physical Review E}, vol.~94, no.~2, p.~022318, 2016.

\bibitem{Kale2017novel}
A.~Kale, ``A novel variable goal appraoch for pedestrian dynamics,'' {\em M.Tech Thesis, IIT Kanpur}, 2017.

\bibitem{mckean2007gender}
K.~A. McKean, S.~C. Landry, C.~L. Hubley-Kozey, M.~J. Dunbar, W.~D. Stanish, and K.~J. Deluzio, ``Gender differences exist in osteoarthritic gait,'' {\em Clinical Biomechanics}, vol.~22, no.~4, pp.~400--409, 2007.

\bibitem{li2008gait}
X.~Li, S.~J. Maybank, S.~Yan, D.~Tao, and D.~Xu, ``Gait components and their application to gender recognition,'' {\em IEEE Transactions on Systems, Man, and Cybernetics, Part C (Applications and Reviews)}, vol.~38, no.~2, pp.~145--155, 2008.

\bibitem{yu2009study}
S.~Yu, T.~Tan, K.~Huang, K.~Jia, and X.~Wu, ``A study on gait-based gender classification,'' {\em IEEE Transactions on image processing}, vol.~18, no.~8, pp.~1905--1910, 2009.

\bibitem{makihara2011gait}
Y.~Makihara, H.~Mannami, and Y.~Yagi, ``Gait analysis of gender and age using a large-scale multi-view gait database,'' in {\em Computer Vision--ACCV 2010: 10th Asian Conference on Computer Vision, Queenstown, New Zealand, November 8-12, 2010, Revised Selected Papers, Part II 10}, pp.~440--451, Springer, 2011.

\bibitem{lv2013two}
W.~Lv, W.-g. Song, J.~Ma, and Z.-m. Fang, ``A two-dimensional optimal velocity model for unidirectional pedestrian flow based on pedestrian's visual hindrance field,'' {\em IEEE Transactions on Intelligent Transportation Systems}, vol.~14, no.~4, pp.~1753--1763, 2013.

\bibitem{moussaid2009experimental}
M.~Moussa{\"\i}d, D.~Helbing, S.~Garnier, A.~Johansson, M.~Combe, and G.~Theraulaz, ``Experimental study of the behavioural mechanisms underlying self-organization in human crowds,'' {\em Proceedings of the Royal Society B: Biological Sciences}, vol.~276, no.~1668, pp.~2755--2762, 2009.

\bibitem{shan2014critical}
X.~Shan, J.~Ye, and X.~Chen, ``Critical walking space requirement for collision avoidance of pedestrians: an experimental study,'' in {\em CICTP 2014: Safe, Smart, and Sustainable Multimodal Transportation Systems}, pp.~2369--2380, 2014.

\bibitem{kramer2021social}
K.~B. Kramer and G.~J. Wang, ``Social distancing slows down steady dynamics in pedestrian flows,'' {\em Physics of Fluids}, vol.~33, no.~10, 2021.

\end{thebibliography}

\pagebreak

\end{document}